\documentclass[12pt]{iopart} 

\usepackage[utf8]{inputenc}
\usepackage[T1]{fontenc} 
\usepackage[english]{babel}
\usepackage[babel]{csquotes} 

\usepackage{multirow} 

\usepackage{graphicx}

\usepackage[separate-uncertainty = true,multi-part-units=single,range-phrase=--,range-units=single]{siunitx}

\usepackage[backend=biber,style=numeric,sorting=none,eprint=false,url=true,giveninits=true]{biblatex}
\newcommand{\um}[1]{\SI{#1}{\micro\meter}}
\newcommand{\mm}[1]{\SI{#1}{\milli\meter}}
\newcommand{\nm}[1]{\SI{#1}{\nano\meter}}
\newcommand{\mbar}[1]{\SI{#1}{\milli\bar}}

\newcommand{\perJperROI}{\si{\per\joule}(\SI{10}{\micro\meter})^{-2}}

\newcommand{\He}{\mathrm{He}} 
\newcommand{\NN}{\mathrm{N}_2}

\newcommand{\revision}[2]{#2}
\newcommand{\revhl}[1]{#1}
\newcommand{\revisionfig}[1]{}

\DeclareUnicodeCharacter{03A7}{\ensuremath{\chi}} 

\begin{document}

\title{Experimental estimates of the photon background in a potential light-by-light scattering study}

\author{
L Doyle$^1$,
P Khademi$^{2,3}$,
P Hilz$^{3,4} $,
A Sävert$^{2,3,4}$,
G Schäfer$^{3,4}$,
J Schreiber$^1$ and
M Zepf$^{2,3,4}$}

\address{$^1$
Fakultät für Physik, Ludwig–Maximilians–Universität München, Am Coulombwall 1, 85748 Garching, Germany}
\address{$^2$
Institute of Optics and Quantum Electronics, Friedrich-Schiller-Universität Jena, Max-Wien-Platz 1, 07743 Jena, Germany}
\address{$^3$
Helmholtz-Institut Jena, Fröbelstieg 3, 07743 Jena, Germany}
\address{$^4$
GSI Helmholtzzentrum für Schwerionenforschung, Planckstraße 1, 64291 Darmstadt, Germany}

\ead{leonard.doyle@physik.lmu.de}

\vspace{10pt}
\begin{indented}
\item[] February 2022
\end{indented}

\begin{abstract}
High power short pulse lasers provide a promising route to study the strong field effects of the quantum vacuum, for example by direct photon-photon scattering in the all-optical regime.
Theoretical predictions based on realistic laser parameters achievable today or in the near future predict scattering of a few photons with colliding Petawatt laser pulses, requiring single photon sensitive detection schemes and very good spatio-temporal filtering and background suppression.
In this article, we present experimental investigations of this photon background by employing only a single high power laser pulse tightly focused in residual gas of a vacuum chamber. The focal region was imaged onto a single-photon sensitive, time gated camera.
As no detectable quantum vacuum signature was expected in our case, the setup allowed for characterization and first mitigation of background contributions.
For the setup employed, scattering off surfaces of imperfect optics dominated below residual gas pressures of \mbar{1e-4}.
Extrapolation of the findings to intensities relevant for photon-photon scattering studies is discussed.
\end{abstract}

%

 

\section{Introduction}

In contrast to the completely empty classical vacuum, quantum physics predict the presence of virtual particle-antiparticle pairs constituting the quantum-mechanical ground state of the vacuum.
These virtual pairs of charged particles act as the mediator of photon-photon-scattering that was proposed by Euler in 1936, effectively making the vacuum a non-linear optical medium \cite{heisenbergFolgerungenAusDiracschen1936,moulinFourwaveInteractionGas1999}.
Early investigations \cite{moulinPhotonphotonElasticScattering1996, bernardSearchStimulatedPhotonphoton2000} revealed first boundaries for the associated scattering cross sections and advised on challenges for photon-photon-scattering experiments in the optical regime.
Other schemes probing e.g. the vacuum birefringence due to external fields or crossing X-ray and laser pulses are being investigated \cite{dellavallePVLASExperimentMeasuring2016, cadeneVacuumMagneticLinear2014, fanOVALExperimentNew2017, heinzlObservationVacuumBirefringence2006, karbsteinProbingVacuumBirefringence2016, karbsteinVacuumBirefringenceXray2021}.
In the high-energy regime, light-by-light scattering could be observed in ultra-peripheral ion collisions involving quasi-real photons \cite{denterriaObservingLightbyLightScattering2013, aadObservationLightbyLightScattering2019, sirunyanEvidenceLightbylightScattering2019}.
To date, no direct light-by-light interaction in pure vacuum has been observed experimentally, but the growing availability of high power laser pulses has refueled the interest in this topic (e.g. \cite{giesAllopticalSignaturesStrongfield2018}) and experimental approaches to find, for example, direct evidence of photon-photon scattering in the optical regime using high intensity laser pulses have recently studied in more detail \cite{lundstromUsingHighPowerLasers2006, kingMatterlessDoubleSlit2010, tommasiniLightLightDiffraction2010, karbsteinEnhancingQuantumVacuum2020, hillProbingQuantumVacuum2017, blinneAllopticalSignaturesQuantum2019}.
In the most simple all-optical scheme, two laser pulses of same frequency $\omega$ collide under an angle of 0 to $\pi$, giving rise to \emph{signal photons} with spectra peaked around the frequencies $\omega$ and $3\omega$, since the lowest order is essentially a four-wave mixing process \cite{moulinFourwaveInteractionGas1999,giesAllopticalSignaturesStrongfield2018, lundstromUsingHighPowerLasers2006}.
In order to maximize chances of detecting scattered photons, one ideally chooses a parameter space (angular, polarization and spectral) that is not occupied by photons from the driving pulses.
The optimization of the ratio of scattered photons over primary photons, often referred to as \emph{discernible} photons, has been addressed in a number of theoretical studies \cite{lundstromUsingHighPowerLasers2006,tommasiniLightLightDiffraction2010, karbsteinBoostingQuantumVacuum2019, klarDetectableOpticalSignatures2020, mosmanVacuumBirefringenceDiffraction2021a}.
While a separation of emission directions and spectra can be achieved for example by frequency doubling one of the beams, employing more than two beams or measuring sum frequencies such as the third harmonic, the interaction of two colliding laser pulses with fundamental frequency remains the simplest experimental configuration and promises highest absolute yields.

It is therefore worthwhile to quantify the challenge posed by minimizing the probability of photons scattering into the phase space by processes other than vacuum photon-photon-scattering events - so called \emph{background photons}.
In practice, the photon distribution of the driving beams is not the only background that needs to be accounted for.
Numerous other sources can contribute, including scattering off rest gas atoms and electrons, surfaces of optics, mechanical elements and walls of the experimental vacuum chamber. Photons from some of these sources may be suppressed by appropriate filtering techniques (spatial, temporal, polarization and, potentially, spectral) which themselves have finite suppression rates.

With this study, we concentrate solely on investigating the background photon signal resulting from a single high intensity, tightly focused laser pulse in the residual gas pressures available with the standard high-vacuum technology commonly in use in high power laser laboratories.
No measurable quantum vacuum signature is expected (see e.g. \cite{karbsteinStimulatedPhotonEmission2015}, a non-zero signal from a single beam requires corrections for large beam cone angles or higher order focus modes.)
and any photons measured constitute the background. We determined the contribution from various sources and scaling with background pressure and draw conclusions on requirements for future measurements with Petawatt laser pulses.

\section{Experiment}
\subsection{Setup and Methodology}
\label{sec:setup_and_methodology}

The experiments were carried out at the JETi-200 laser facility at the Helmholtz Institute Jena, with a short pulse Ti:Sapphire system capable of peak powers in excess of 200TW.
In this campaign we are focusing predominantly on photon background originating from linear scattering. Consequently the laser was operated at reduced energy, providing approximately \SI{1}{\joule/\second} (repetition rate of 5 Hz with \SI{175}{\milli\joule} per pulse) in focus at \SI{24}{\femto\second} full-width-at-half-maximum (\emph{FWHM}) pulse duration (with a peak power of \revision{Number corrected, simple calculation error}{$\approx\SI{7}{\tera\watt}$}) at a wavelength of \nm{800}.

\begin{figure}
    \begin{center}
	    \includegraphics[width=.48\textwidth]{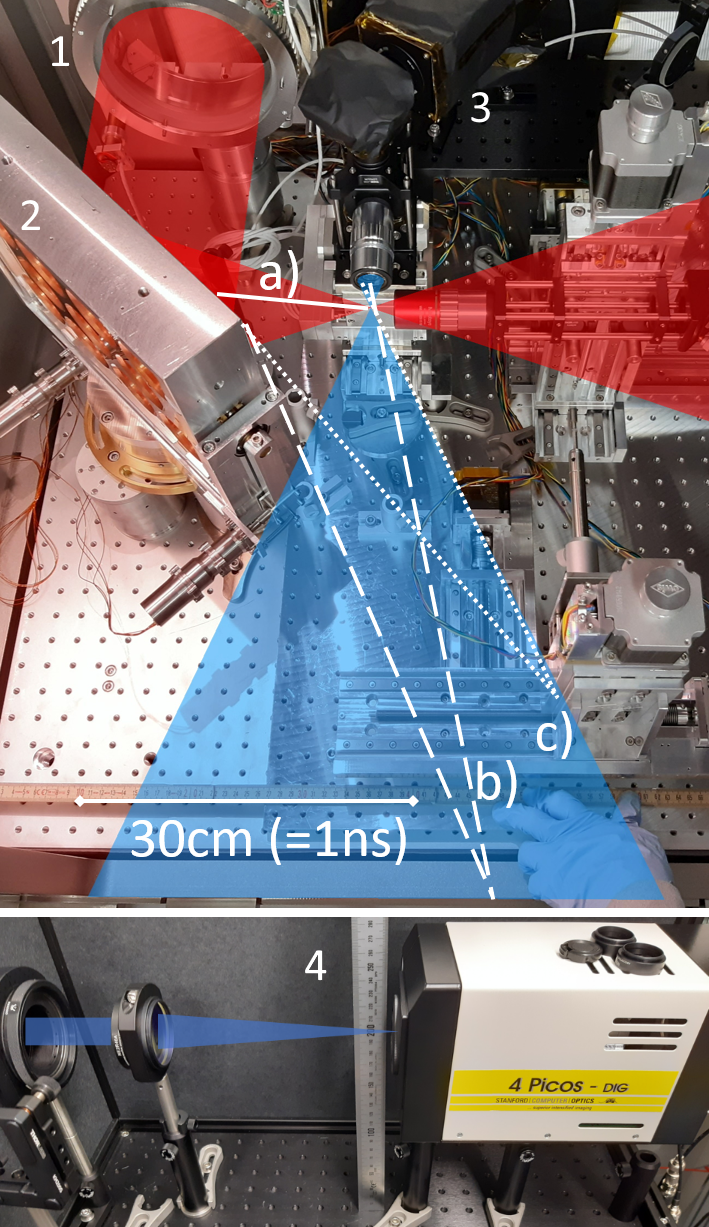}
	\end{center}
	\caption[Setup picture]{\label{fig:setup_picture}Setup in vacuum chamber \textbf{(top)} and on air side \textbf{(bottom)}. \textbf{1} $\lambda/2$-plate (only present for vertical polarization), \textbf{2} OAP, \textbf{3} collection optic and light-tight guiding and \textbf{4} 4Picos single-photon camera. Light paths and equivalent time to focal region \textbf{a)} for main pulse ($\approx\SI{0.5}{\nano\second}$), \textbf{b)} nearest chamber wall (inside viewing cone, $\approx\SI{4}{\nano\second}$) and \textbf{c)} for nearest object outside of, but close to viewing cone ($\approx\SI{3}{\nano\second}$).}
\end{figure}

\begin{figure}
    \begin{center}
    	\includegraphics[width=.4\textwidth]{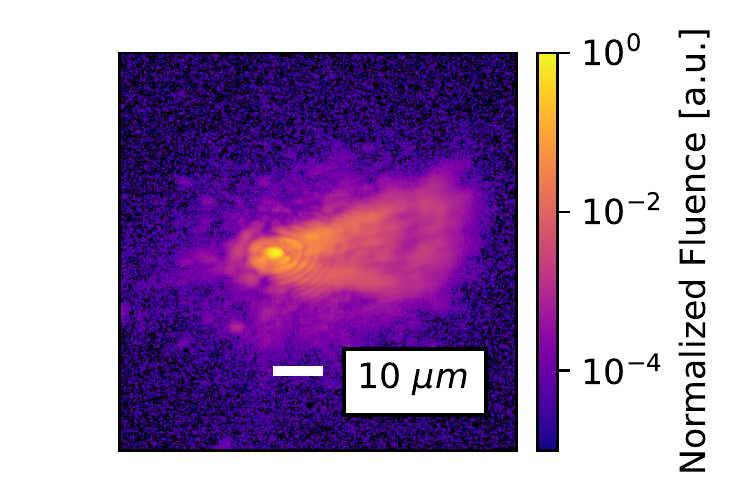}
    \end{center}
	\caption[HDR Focus]{\label{fig:focus} Spatial focus distribution with high dynamic range (HDR) (logarithmic scale). 34\% energy inside $1/\mathrm{e}^2$ area surrounded by pedestal of $>\um{30}$ vertical extent (as observed by collection optic) at >1\% peak fluence. 76\% of total energy in area above 1\% peak fluence. Close to diffraction limited performance via adaptive optics is achievable at the JETi-200 for future studies.}
\end{figure}

The beam was focused using a \ang{90} off-axis parabolic mirror (\emph{OAP}) with f/1.5 focusing, as shown in \fref{fig:setup_picture}, thus providing a wide focusing cone angle similar to that of a photon-scattering experiment.
As achieving a high peak intensity was not required, the FWHM focal spot diameter of \um{2.2} achieved without compensating for residual phase errors introduced by the parabola was deemed satisfactory.
\Fref{fig:focus} shows the spatial focus distribution, the central spot contained 34\% of the pulse energy inside the $1/\mathrm{e}^2$ radius with a low intensity pedestal at above 1\% peak intensity containing 76\% of the pulse energy and extending $>\um{30}$ in the vertical and $>\um{50}$ in the horizontal directions, respectively.
\revision{Added number for peak intensity}{Overall this leads to an estimated peak intensity of {\SI{5e19}{\watt\per\centi\meter\squared}}}.
\revision{Details about energy measurement}{The pulse energy in focus was determined by integrating the beam brightness on calibrated beam monitor cameras in the laser chain and subsequently corrected for the measured transmission of remaining optics. The uncertainty of the calibration is estimated to be less than the typical shot-to-shot fluctuations of $<2\%$ rms.}
We note that the diamond machined copper OAP \revision{Added manufacturer details}{({\SI{4.4}{\nano\meter}} surface roughness, Kugler, Germany)} scatters $\approx 0.2\%$ of the laser light uniformly in $2\pi$, \revhl{calculated via the} \enquote{total integrated scatter}\revision{Refined estimate, added reference}{ due to surface roughness as described in} \cite{stoverOpticalScatteringMeasurement2012}, affecting scattering from walls or other solid objects into the beam path.
The light scattered from the focal region was collected under an angle of \ang{90} in the horizontal plane using a long working distance microscope objective ($NA=0.40$, \enquote{M Plan Apo NIR B $20\times$}, Mitutoyo, Japan).
The polarization of the laser could be changed from horizontal (along observation axis) to vertical (perpendicular to observation axis) by introducing a $\lambda/2$-plate \revision{Added manufacturer details}{(uncoated mica, B Halle Nachfl., Germany)} before the OAP.
\revision{Added transmission loss of L/2-Plate}{As the data recorded at vertical polarization is more relevant in the following, the pulse energy of {\SI{175}{\milli\joule}} is stated for this case. At horizontal polarization, without the $\lambda/2$-plate, the energy was increased by $\approx11\%$ (calculated from the refractive index of the uncoated mica plate).}
The collected light was guided by a light shielded tube system outside of the vacuum chamber, where an achromatic lens (\enquote{AC508-200-B-ML}, Thorlabs, US) formed a $20\times$ magnified image of the focal region on the primary diagnostic, a temporally gated, single-photon sensitive image-intensified CCD camera (4Picos, Stanford Computer Optics, US).
Considering the spectral characteristics of the optics and camera involved, photons in a spectral range of \SIrange{400}{900}{\nano\meter} were detectable.
The overall detection efficiency for light at the fundamental wavelength of \nm{800} is estimated to be 9\%, while around \nm{400} wavelength (second harmonic) it is reduced to $\approx 1\%$.

Special care was taken to keep the direct viewing cone free of any scattering objects, since light scattered into this cone would be efficiently transported onto the camera.
Note that light scattered from objects outside the viewing cone may still reach the camera through scattering in the collection optic and imaging system, albeit at reduced efficiency.
By introducing $\He$ gas into the entire chamber, the residual gas pressure could be tuned from \SI{7e-6}{\milli\bar} to \revision{Many occasions: Number changed due to pressure correction}{{\SI{1e-3}{\milli\bar}}}.
Additionally, some data could also be recorded at pressures up to \SI{1}{\milli\bar} while operating the laser during pumpdown of the vacuum chamber.
\revision{Added details.}{The pressure was determined via a combined Pirani and cold-cathode gauge ($\pm30\%$ precision, $\pm5\%$ repeatability, PKR251, Pfeiffer Vacuum, Germany) situated $<\SI{1}{\metre}$ from the focus point.
When {$\He$} is used, a correction factor must be applied to the pressure reading. To account for the background pressure of $>\mbar{7e-6}$ of residual air always present, the correction was applied assuming independent partial contributions to the total pressure by {$\He$} and residual air.
Since they may not be truly independent in reality, the error bars on any data corrected were extended to include also the uncorrected value.}
During pumpdown, a glass window was closed, separating the experimental chamber from the evacuated beamline, leading to an overall \revision{Quantified fluence redcution}{reduction of peak fluence by 70\% compared to that shown in} \fref{fig:focus} (due to 15\% energy loss by reflection from the uncoated window and an increased focal spot size).

\begin{figure}
    \begin{center}
	    \includegraphics[width=.48\textwidth]{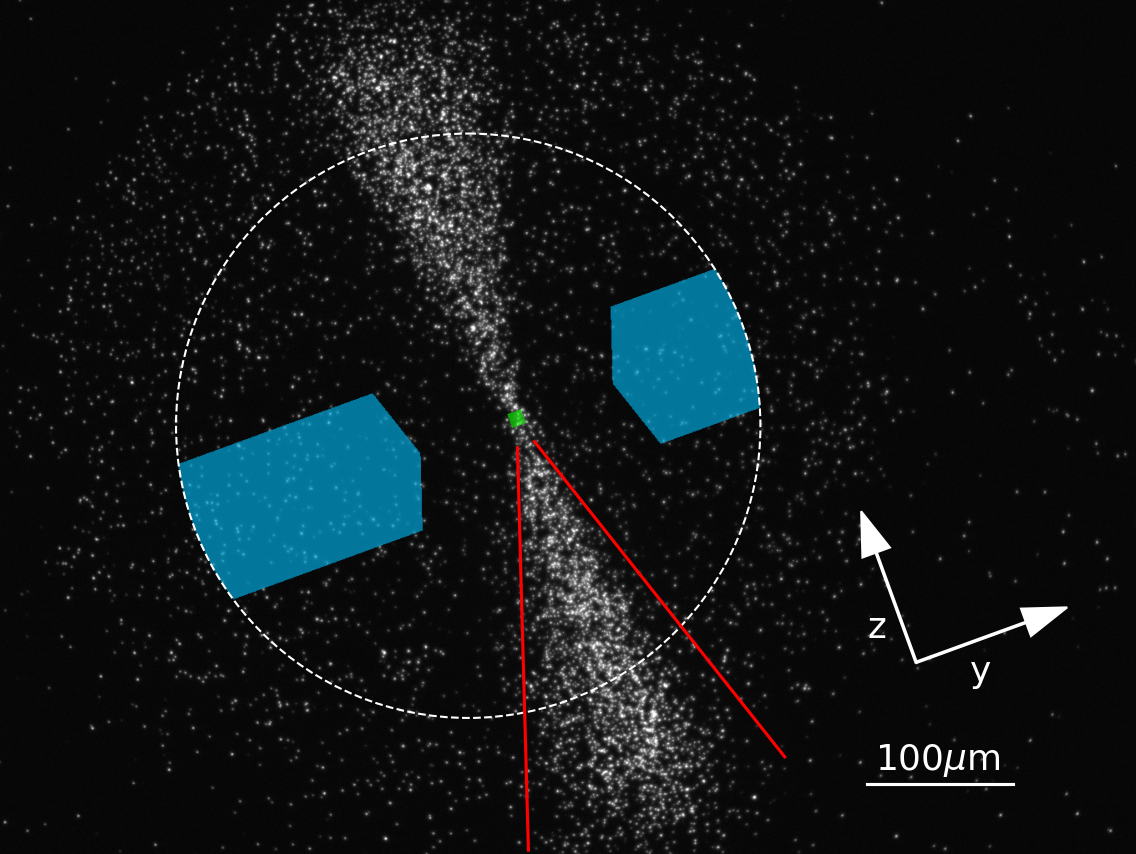} 
	\end{center}
	\caption[Example shot]{\label{fig:exampledata}\revisionfig{Improved details.} Example shot of a single photon image at $p=\SI{1.3e-3}{\milli\bar}$ and vertical polarization. The laser cone (\emph{bow tie}) is clearly visible, but surrounded by a homogeneous signal only limited by vignetting toward the image edges. \textbf{Green:} focus ROI of $10\times\SI{10}{\micro\meter\squared}$, \textbf{blue:} out-of-cone ROI with margin to beam cone and clipped to vignette (dashed circle). \textbf{Red:} nominal beam cone by f-number. \textbf{Arrows:} $z$ is laser propagation direction, $y$ points upward in experiment. The rotation is due to the imaging setup. }
\end{figure}

The gating window of the 4Picos camera was set to \SI{1}{\nano\second} and the delay adjusted such that the camera recorded photons propagating through the focal region simultaneously with the main pulse.
We define this time as $t\equiv0$.
\revision{added explanation}{The correct delay was determined by scattering attenuated pulses off a glass needle placed in focus}.
The timing jitter of the trigger with respect to the main pulse was measured to be below \SI{100}{\pico\second} (rms).
\Fref{fig:exampledata} shows an exemplary image taken at a pressure of \mbar{1.3e-3} and vertical polarization, clearly indicating the shape of the laser cone and focus as viewed from the side, resembling a \emph{bow tie} shape.
Since no other scattering objects are close, the image is formed primarily by scattering off free electrons as discussed below.
Close to the image edges, transmission efficiency is reduced by vignetting and consequently analysis concentrates on the central part of the image.

Inside the non-vignetted area (inside dashed circle in \fref{fig:exampledata}), two regions of interest (\emph{ROI}) are compared.
Firstly, a square box corresponding to an area of $10\times\SI{10}{\micro\meter\squared}$ around the waist of the beam is defined, large enough to capture all photons originating from the focal volume.
For a vacuum photon scattering experiment, the source region for signal photons can be estimated as the intersection of all pulses' focal volumes, approximated by cylinders with focal diameter $d_{FWHM}$ and length of the Rayleigh range $z_R$.
Outside this region, the spatio-temporal overlap of the colliding pulses is negligible\cite{giesAllopticalSignaturesStrongfield2018}.
\revision{improved clarity}{In addition to the bow tie, a homogeneous scattering signal covering the entire non-vignetted image is observed. To study this, a second ROI outside the laser cone is defined as the \emph{out-of-cone} reference,}
highlighted in \fref{fig:exampledata}.
The number of recorded photons in this out-of-cone area is scaled down to \SI{100}{\micro\meter\squared} to match the focal ROI.

\begin{figure}
    \begin{center}
	    \includegraphics[width=.88\textwidth]{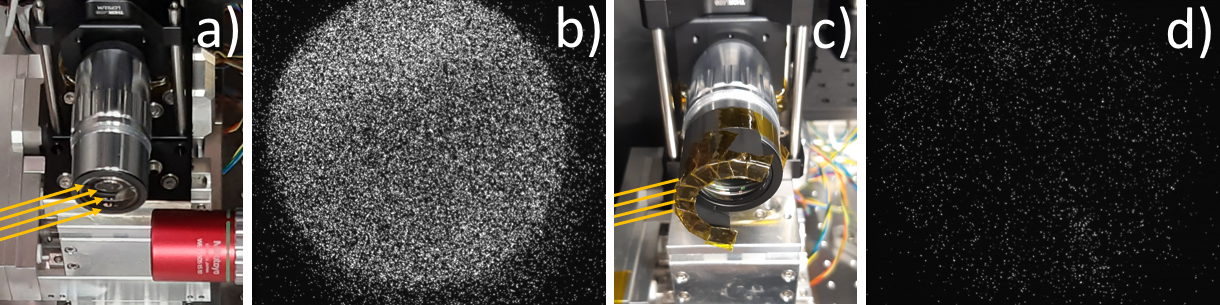}
	\end{center}
	\caption[Light shielding baffle]{\label{fig:baffleinstall}Objective before \textbf{(a)} and after \textbf{(c)} covering with a light shielding baffle, stray light rays from OAP sketched. (Focus diagnostic objective on right removed during full power shot). Resulting raw images \textbf{(b)} without any baffle at \mbar{1e-5} and \textbf{(d)} for same settings with baffle, showing significantly reduced background.}
\end{figure}

\revision{shortened}{In a first step, this homogeneous background could be significantly reduced by introducing a simple light shielding baffle, restricting the line of sight from the OAP to the collection optic aperture.}
This is highlighted in \fref{fig:baffleinstall}, comparing the background signal in the initial setup and after a simple light shielding baffle was installed.
All other data shown was recorded with the baffle installed.

To estimate the actual number of photons collected from a certain region by the collection optic, the camera response was calibrated in a dedicated setup.
For a fixed gain of the image intensifier, the mean pixel count associated with a single photon hit was determined by integrating the brightness in a low-hit-count region and dividing by the number of hits determined with a peak finding algorithm.
For non-saturated images, the total number of photons counted in a certain region is therefore the summed pixel brightness in that region divided by the mean brightness of a single photon detection event.
In contrast to peak-finding, this method is also applicable at lower gain where the images do not show single hits. The drawback is an increased sensitivity to camera dark noise, since no thresholding is applied. This statistic uncertainty is added to the error bars where applicable.
The number of photons on the detector were finally converted to the actual number of photons collected by accounting for the quantum efficiency of the detector (16.6\% at \nm{800} central laser wavelength) and the spectral transmission characteristics of the collection optic setup.

The scattering signal was studied for a range of trigger delays, rest gas pressures, laser energies and two polarizations .
For every set of parameters chosen, several shots
were recorded.

\revisionfig{Changed order of sections: First pressure dependence is clarified for inside/outside bow tie, then the brightness reduction due to ponderomotive cavitation is discussed. The static scattering sources are discussed last, including the time scan.}

\subsection{Scattering from residual gas}
\label{sec:pressure_scatter}
\begin{figure}
    \begin{center}
	    \includegraphics[width=.9\textwidth]{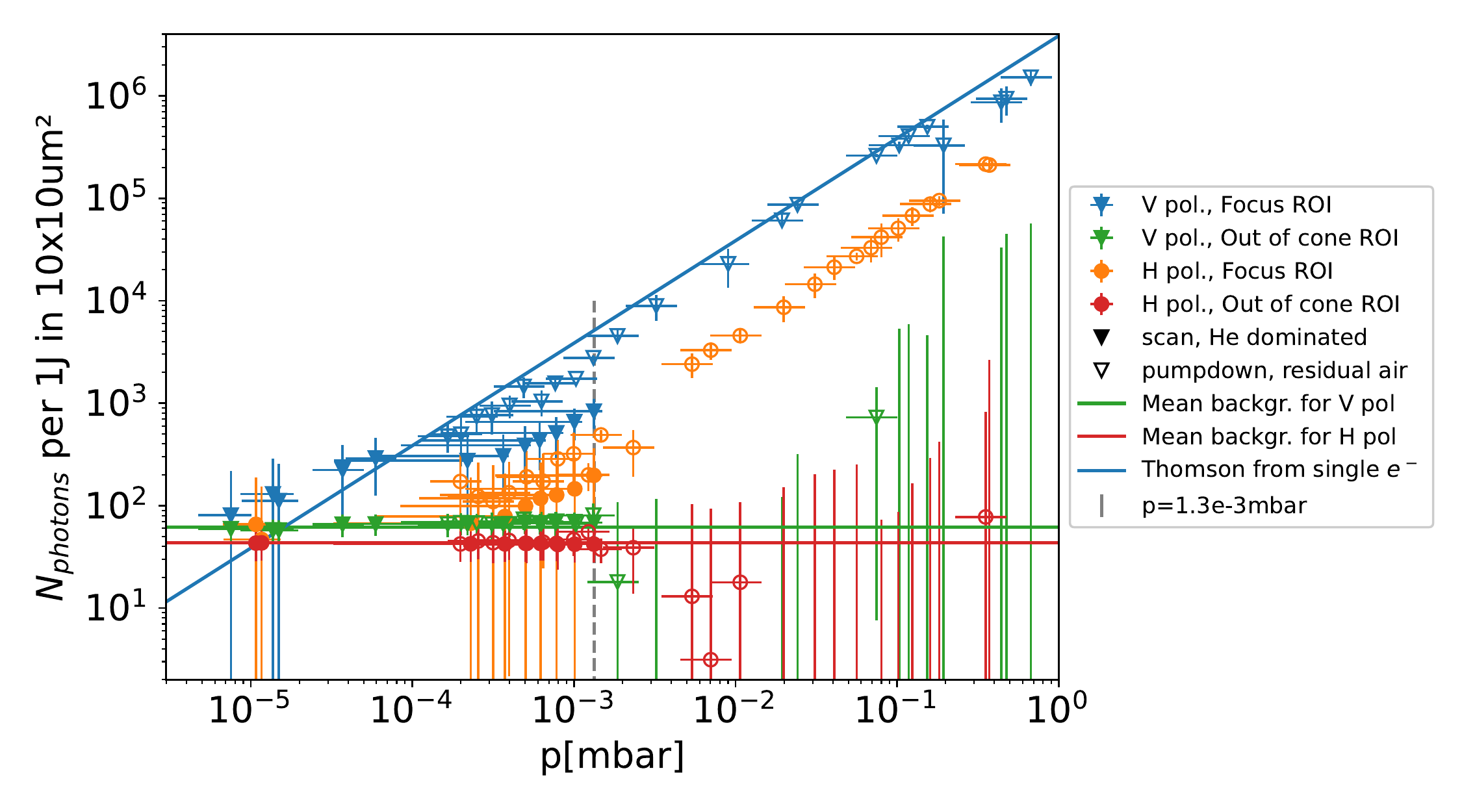}
	\end{center}
	\caption[Pressure dependence]{\label{fig:pscan}
	\revisionfig{Added error bars, applied pressure correction, tidied up legend, moved identified contributions to figure 8, improved caption.}
	Dependence of the scattered signal on the residual gas pressure for both polarizations inside the focal ROI and area-equivalent for the out-of-cone ROI. Scaled to \SI{1}{\joule} laser energy for comparability. Horizontal error bars: measurement uncertainty and $\He$-correction where applicable. Vertical error bars: standard deviation, at low counts dominated by camera dark noise. Vertical bar at \mbar{1.3e-3}: pressure at which \fref{fig:lineout} and \ref{fig:tscan} were evaluated. Blue line: Thomson scattering estimate from a free electron density equal to the rest gas particle density.}
\end{figure}

\revisionfig{Moved pressure-dependence plot first, integrated scatterer density below.}
\Fref{fig:pscan} shows the pressure dependence of the number of photons registered per \SI{1}{\joule} of laser energy in $10\times\SI{10}{\micro\meter\squared}$ inside the focal ROI and in the out-of-cone ROI in a time window of \SI{1}{\nano\second} around $t=0$.
The number of photons was normalized to \SI{1}{\joule} to correct for the energy differences due to the glass shutter and $\lambda/2$-plate where relevant.
During pumpdown of the chamber, the pressure dropped continuously from atmospheric pressure.
After the final pressure was reached and the glass shutter was opened, $\He$ gas was introduced to increase the pressure again.
For both horizontal and vertical polarization, the signal inside the focus region increases approximately linearly with pressure, while the out-of-cone contribution stays constant also for higher pressures.
This indicates that outside the nominal beam cone, scattering from the rest gas is negligible compared to other, pressure-independent scattering sources.
Inside the focal ROI, this contribution adds a constant offset to the number of photons measured.
Below $\approx\mbar{1e-4}$, the amount of scattered light in focus and outside the laser cone each converge to the same value for each polarization, namely $(62\pm7)$ photons per \SI{1}{\joule} per $10\times\SI{10}{\micro\meter\squared}$ for vertical and $(43\pm4) \perJperROI$ \revision{Number changed due to L/2-plate energy rescaling}{photons} for horizontal polarization.

\revisionfig{Moved paragraph here from discussion and reworded for clarity.}
The approximately linear dependence of photon number on gas pressure is related to Thomson scattering of the laser pulse by free electrons. Field ionization dominates already at intensities beyond $10^{15}\si{\watt\per\centi\meter\squared}$\cite{changClosedformSolutionsProduction1993}. One hence expects multiply ionised residual gas atoms and correspondingly high electron density.
This is in line with the observed difference of total scattering for the same number density of gas particles for $\He$ or residual air dominated data points. In the former case, an electron density up to twice the gas particle density $n_0$ can be expected, while in the latter case e.g. $\NN$ or $\mathrm{O}_2$ can contribute several electrons each.
The blue solid line in \fref{fig:pscan} represents the Thomson scattering signal from an electron density that equals the gas particle density at the respective pressure, corresponding to single ionization.
This yield was calculated from integrating the emission pattern of a dipole directed perpendicular to the observation axis over the acceptance angle of the collecting objective.
Both for $\He$-dominated as well as for residual air dominated data points the signal in focus for vertical polarization lies below this estimate and indicates that the electron density is significantly reduced within the focal spot volume. This will be analysed in more detail below.
The same simple calculation at horizontal polarization (where the fictive dipole points perfectly into the direction of observation of the objective) predicts a reduction of the signal by a factor of 12, while the experimental data (orange symbols in \fref{fig:pscan}) shows a ratio of \revision{Number refined due to energy rescaling}{$\approx6\times$} between horizontal and vertical polarization.
This discrepancy can be expected since the short focal length of the OAP leads to dipole orientations that deviate considerably from the assumption.
The scattering cross sections for bound atoms and electrons (Rayleigh scattering) and charged nuclei are orders of magnitude lower at the relevant wavelengths and can be neglected.

\revision{Reworded entire paragraph about figure 6 for clarity}{
In addition to the analysis of the focal region, the effective scatterer density along the laser propagation axis was determined for groups of shots at similar settings.}
\begin{figure}
    \begin{center}
	    \includegraphics[width=\textwidth]{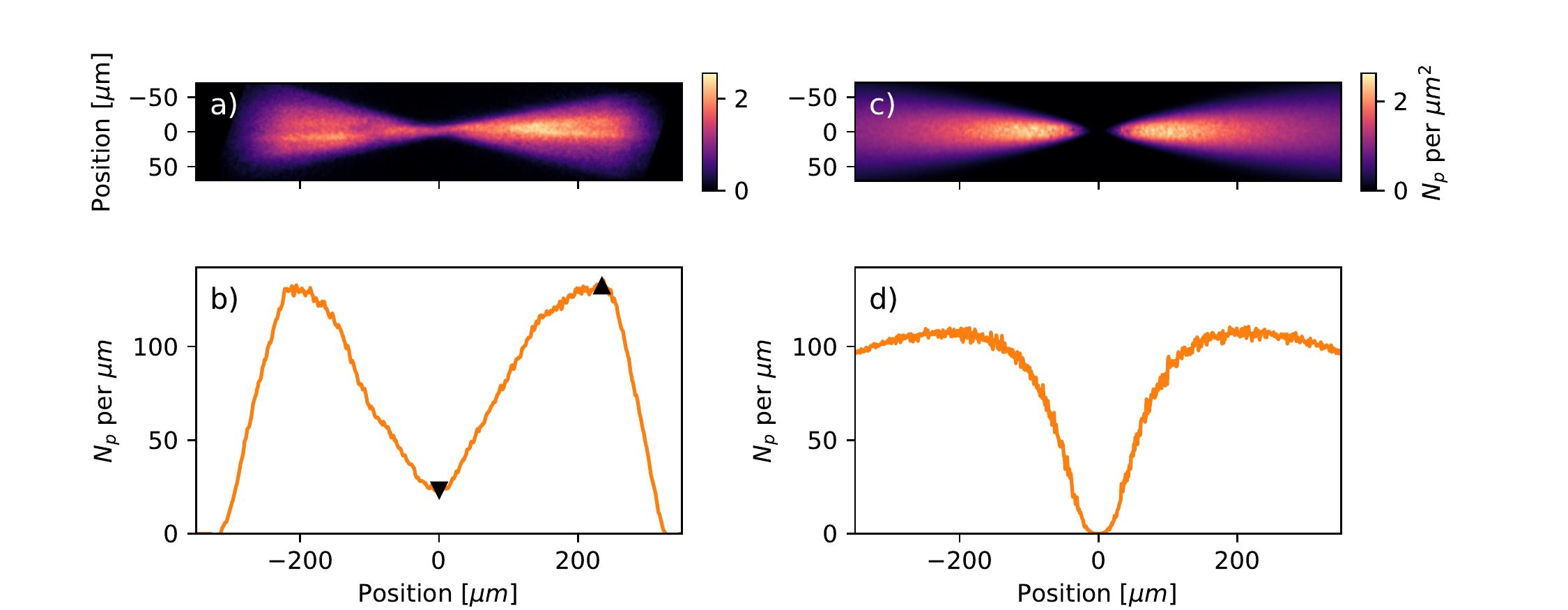}
	\end{center}
	\caption[Beam lineout]{\label{fig:lineout}
	\revisionfig{Tidied up a) and b), improved caption, added c) and d) for numerical model.} 
	\textbf{a)} Average number of photons per shot $N_p$ per \si{\micro\meter\squared} over 242 shots at ${\approx\mbar{1.3e-3}}$ and \SI{175}{\milli\joule} pulse energy each.
	The out-of-cone background was subtracted. \textbf{b)} Image a) integrated along vertical extent ($\pm\SI{70}{\micro\meter}$), giving number of photons per \si{\micro\meter} slice along the optical axis. A reduction of integrated intensity is clearly visible in the center region. For the image shown, the reduction factor between minimum in focus and maximum peak is $5.6\times$ (black triangles). Laser propagation from right to left.
	\textbf{c)} Numerical simulation of Thomson scattering by electrons ionized from $\He$ by an ideal Gaussian laser pulse. Details see text. 
	\textbf{d)} Image c) integrated along vertical extent ($\pm\SI{70}{\micro\meter}$), giving number of photons per \si{\micro\meter} slice along the optical axis. The reduction of integrated intensity in focus is even more significant compared to b), most likely due to ideal focusing of the simulated Gaussian.
	}
\end{figure}
\Fref{fig:lineout}a shows the averaged number of detected photons per \si{\micro\meter\squared} per shot over 242 shots recorded at \SIrange{1e-3}{1.5e-3}{\milli\bar} and \SI{175}{\milli\joule} at vertical polarization.
The signal level determined in the out-of-cone ROI was subtracted from each image before averaging.
\Fref{fig:lineout}b shows the integral of this image along the entire vertical extent and yields the total number of photons scattered per \si{\micro\meter} slice along the optical axis. The scatter yield drops significantly towards the focus by a reduction factor of $5.6$ compared to the maximum at $>\SI{200}{\micro\meter}$ ($>46$ Rayleigh lengths) upstream.
More than \SI{250}{\micro\meter} away from focus, the integrated intensity drops at the image edges due to vignetting.

\revisionfig{Added new section about theoretical model [Gibbon] and numerical model.}
The observed reduction of scattered photons in the volume of highest laser intensity with increasing laser intensity is expected. Ponderomotive cavitation occurs for \cite{gibbonShortPulseLaser2005} $I_{18}\lambda_{\mu}^2>1/20 n_{18}\sigma_{\mu}^2$ ($I_{18}$ intensity in $10^{18}
\si{\watt\per\centi\meter\squared}$, $n_{18}$ electron density in $10^{18} cm^{-3}$, $\lambda_{\mu}$ wavelength in \si{\micro\meter} and $\sigma_{\mu}$ the $1/e$ radius in intensity in \si{\micro\meter}). In the above example for \mbar{1.3e-3} the electron density of singly ionized Helium is $n_{18}=\num{3e-5}$ and the peak intensity in focus is $I_{18}=50$, hence, we exceed this threshold by far.
Assuming a perfect Gaussian pulse with $\sigma_{\mu}=0.7$ (beam waist $w_0=\sqrt{2} \sigma=\SI{1.05}{\micro\meter})$, cavitation on axis is expected as long as 
$I_{18}\frac{\sigma_0^2}{\sigma(z)^2}\lambda_\mu^2 > 1/20 n_{18}\sigma(z)^2$,
that is for $z<(20 I_{18} \lambda_\mu^2/(\sigma_{\mu}^2 n_{18}))^{1⁄4} z_R\approx 74 $ times the Rayleigh length $z_R$ or \SI{320}{\micro\meter}, larger than the field of view of our microscope.

On the other hand, the laser pulse duration $\sigma_t=\tau_{FWHM}/\sqrt{4\mathrm{ln}2}$ is too short for establishing an equilibrium situation on which the above estimate is based, that is the plasma period $1.1ps/\sqrt{n_{18}}\gg \sigma_t$. This also means that the force of the ionic background on the electrons can be ignored for early times of electron motion.
We therefore can attempt to explain the photon scatter image and vertical integral of \fref{fig:lineout}a and b by a simple model.
$V\cdot n_0$ Helium atoms are randomly distributed within a predefined simulation volume $V=\Delta x \Delta y \Delta z$, where $n_0=\SI{3.2e13}{\per\cubic\centi\meter}$ is the particle (atom) density corresponding to a residual gas pressure of \mbar{1.3e-3}.
We consider a Gaussian laser intensity distribution $I_L(t,r,z) = I_0 w_0^2/w(z)^2 \cdot \exp[-2 r^2/w(z)^2-(ct-z)^2/(c\sigma_t)^2]$, where $r=(x^2+y^2)^{1/2}$ and $w(z) =w_0 ( 1+(z/z_R)^2 )^{1/2}$ with $w_0=\SI{1.05}{\micro\meter}$, $z_R=\SI{4.3}{\micro\meter}$, $\sigma_t=\tau_{FWHM}/\sqrt{4\mathrm{ln}2}=\SI{14.4}{\femto\second}$ and $I_0=\SI{4e20}{\watt\per\cm\squared}$, corresponding to an ideal Gaussian pulse with same $d_{FWHM}$ and $I_{peak}$ as the theoretical, diffraction limited best focus in the experiment. Electrons are considered free when the intensity exceeds the 50\% threshold intensity for field ionization, $I_k$ \cite{changClosedformSolutionsProduction1993}.
The electron $i$ will hence start moving in the laser field for times $t>t_{0i}=z_i/c - \sigma_t (\ln[(I_0/I_k)(w_0^2/w^2(z_i)]-2r^2/w(z_i)^2)^{1/2}$.
We then solve the equation of motion for each electron, considering only the transverse ponderomotive force, $\dot{\vec{p}}_i(t)=-1/(2cn_c) \nabla_\perp I(t,r_i(t),z_i)$ and $\dot{\vec{r}}_i(t)=c\vec{p}_i(t)/((mc)^2+p_i^2(t))^{1/2}$, where $n_c=\SI{1.74e21}{\per\cubic\cm}$ is the critical density and $z_i$ the initial (random) position of the electron along laser propagation direction. The numerical solution yields the trajectories $\vec{r}_i(t)=(x_i(t),y_i(t),z_i)$. We then calculate for each electron the number of photons that it scatters via Thomson scattering while moving along its trajectory, $\Delta N_i(t,x_i,z_i)=\sigma_T/(2\hbar\omega_L)(I_L(t,x_i(t),z_i)+I_L(t+\Delta t,x_i(t+\Delta t),z_i)) \Delta t$. This contribution is distributed onto a predefined $(x_k,z_l)$ grid. Finally, all contributions in a single cell $x_k,z_l$ are summed to yield a calculated side view image of the focal volume (\fref{fig:lineout}c) and the corresponding integrated yield along propagation direction (\fref{fig:lineout}d). The reduction is clearly visible and stronger than in the experimental case. This is likely due to the fact that the pulse in the experiment is not a perfect Gaussian.
Overall, this simple model is in good agreement with the measured effective scatterer density considering the idealized assumptions made.

We do not observe the complete extent of the volume that is affected by this dynamic behavior of the scattering electrons. Nevertheless it is interesting to state that the reduction was observed for a variety of other experimental settings. \Tref{tab:lineoutfactor} summarizes the reduction factors between the focus brightness and maximum brightness observed in each graph for the cases that we analysed. For highest pulse energies, a reduction factor of 4.6 to 10.5 was determined for both polarizations at best focusing.
For pressures $>\mbar{2e-3}$, the reduction factor was limited to $1.7-4.3\times$, likely due to the degraded focus with the transmissive glass shutter present.
The reduction is also moderated to $2.2\times$ ($1.7\times$) for \SI{20}{\milli\joule} (\SI{4}{\milli\joule}) of pulse energy, respectively.
This trend supports the importance of the ponderomotive cavitation with increasing intensity once more.

\begin{table}
	\caption{\label{tab:lineoutfactor}
	\revisionfig{Improved format, corrected energies for H-pol (due to L/2-plate).}
	Summary of scattering reduction factors as exemplified in \fref{fig:lineout} for other parameters.
	Note that the peak fluence was reduced when glass shutter was closed and increased for horizontal polarization, where the $\lambda/2$-plate was removed. (**) Signal low.}
	\begin{indented}
		\item[]\begin{tabular}{@{}l|llllll}
			\br
			Glass shutter & Closed & Closed & Open & Open & Open & Open \\
			p[mbar] & $>\num{2e-3}$ & $\num{1.3e-3}$ & $\num{1.3e-3}$
				& $\num{3e-4}$ & $\num{1.3e-3}$ & $\num{1.3e-3}$ \\
			\mr
			E[mJ] for V pol. & 148 & 148 & 175 & 175 & 20 & 4 \\
			V. polarization & 3.7 & 4.3 & 5.6 & 10.5 & 2.2 & 1.7 \\
			\mr
			E[mJ] for H pol. & 164 & 164 & 194 & 194 & 22 & 4 \\
			H. polarization & 1.7 & 3.3 & 4.6 & ** & ** & ** \\
			\br
		\end{tabular}
	\end{indented}
\end{table}

We stress that the model contains several simplifications that might affect the results. In particular, although the equation of motion is relativistically correct, we ignore the force in laser propagation direction. We also do not resolve the oscillation period and hence have no measure of describing coherent effects or scattering in frequency ranges other than covered by the laser. Nevertheless, because in our experiments the signal recorded was dominated by light at the fundamental frequency (this was tested by inserting a broadband dielectric mirror optimized for reflection of light around \nm{800} wavelength before the camera), we are confident that the effect of ponderomotive cavitation in sub-relativistic intensity regions dominates the reduction of Thomson scatter background.

\subsection{Scattering from static scatterers}
\label{sec:static_scatter}
\revisionfig{New subsection for static scattering discussion, including time scan.}
Toward a potential quantum vacuum experiment, in addition to the scattering by the rest gas it is also crucial to understand and subsequently minimize the static, pressure-independent scattering sources.
To study this contribution further, the time dependence of the signal can provide some additional insight.

\begin{figure}
    \begin{center}
	    \includegraphics[width=.6\textwidth]{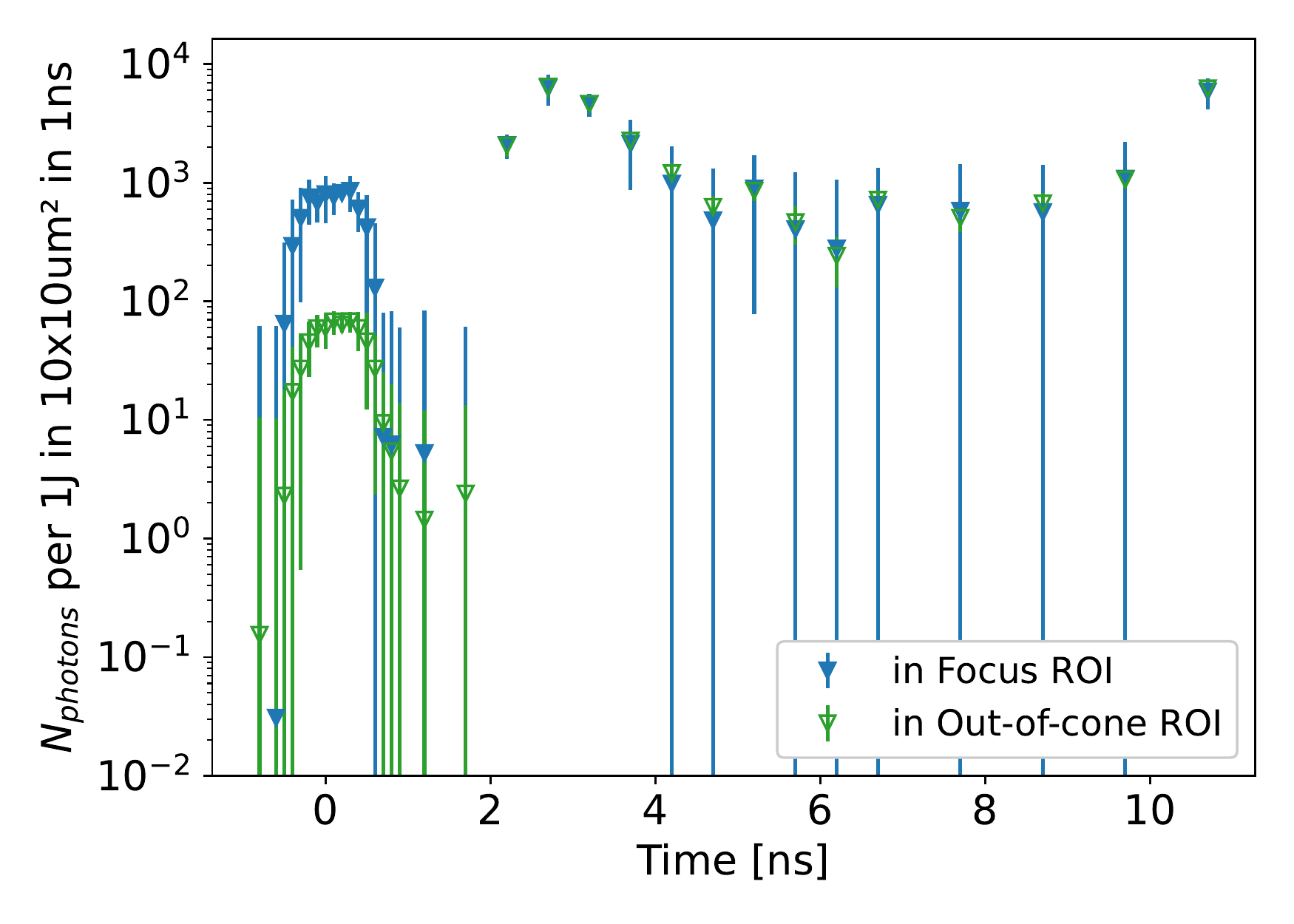}
	\end{center}
	\caption[Time scan]{\label{fig:tscan}Time delay scan. Number of photons recorded in a \SI{1}{\nano\second} gate width in the focus ROI and area-equivalent from out-of-cone ROI for vertical polarization at \mbar{1.3e-3}, per \SI{1}{\joule} of pulse energy. Error bars show standard deviation. The signal at $t=0$ is equal to that shown in pressure scan. Due to log-scale, data points $\le0$ not visible.}
\end{figure}

By varying the delay between the laser pulse and the camera exposure trigger, a temporal scan of the photon signal was obtained.
\Fref{fig:tscan} shows a time scan of the photon number in a \SI{1}{\nano\second} window recorded at $p=\mbar{1.3e-3}$ for vertical polarization, again scaled to \SI{1}{\joule} of laser pulse energy.
The number of photons in the focus ROI around $t=0$ is $(800\pm300) \perJperROI$ and is equal to that shown in \fref{fig:pscan} \revision{added details}{(uncertainty due to photon statistics and camera dark noise)}.
Notably, the signal inside the focal ROI and the out-of-cone region only differ around $t=0$, i.e. all the photons registered at other times must have a pressure-independent origin.
The temporal width of this first peak is determined by the convolution of the laser pulse temporal shape ($\tau_{FWHM}\approx \SI{24}{\femto\second}$) and the gating function of the camera ($\tau_{exposure}=\SI{1}{\nano\second}$), thus yielding a $\approx\SI{1}{\nano\second}$ peak.
After $t>\SI{1}{\nano\second}$, the signal in focus drops again to less than one photon per Joule, indicating very few photons crossing the focal region after the main pulse passed.
The signal next rises at $t>\SI{2.5}{\nano\second}$ where $(6300\pm1300) \perJperROI$ photons are recorded in \SI{1}{\nano\second}.
This temporal delay matches the additional path length travelled by stray light of the main pulse bouncing off objects located in the experiment chamber as shown by path c) in \fref{fig:setup_picture}.
Between $t=\SI{5}{\nano\second}$ and $t=\SI{10}{\nano\second}$, the signal decays but does not drop to zero again, which is in accordance with temporally smeared out scattering from the large walls with a minimum path delay highlighted as b) in \fref{fig:setup_picture}.
Finally, at $t>\SI{10}{\nano\second}$ the signal increases drastically and the detector saturates (even with lower gain applied).
This temporal delay matches the additional path length travelled by the main pulse bouncing off the (bare aluminium) chamber walls and back to the focal region, giving rise to a signal several orders of magnitude larger than the signal at $t=0$.

\revisionfig{Moved paragraphs here from discussion.}
This behaviour can be understood as follows.
The laser pulse carries the majority of its energy in a time window much shorter than the \SI{1}{\nano\second} gate width around the main peak.
Most of this light travels inside the cone expected from geometrical optics.
Diffraction effects can be neglected, except close to the focal region ($\ll\mm{1}$).
This major part of the pulse can only contribute light around $t=0$ via scattering directly from free electrons and residual gas atoms or molecules.
Further downstream, photons from the main beam hit the chamber walls, get scattered in all directions and can enter the imaging system after a minimum round trip of $\approx\SI{3.3}{m}$, giving rise to the large signal after \SI{11}{\nano\second}.
One possibility for additional background signal around $t=0$ therefore arises by light travelling $\approx\SI{11}{\nano\second}$ ahead of the main pulse.
Although high power laser pulses are often accompanied by pre-pulses and \emph{amplified spontaneous emission} (ASE) on the picosecond and nanosecond time scale, a contribution over \SI{11}{\nano\second} ahead is assumed to be negligible, supported by the observation that immediately before $t=0$ (corresponding to ASE at \SI{-12}{\nano\second} ahead of the main pulse) the background signal is below detection threshold. A possible contribution of ASE at shorter timescales ahead of $t=0$ will be discussed below.

Apart from the light travelling inside the expected path, a small but relevant portion of the energy will be scattered by imperfect surfaces of the beam transport optics.
With hardly any delay, photons escaping the main cone can potentially enter the collection optic at a large angle, where they further scatter in the lens media and off the tube walls.
The importance of this contribution is evidenced by the images shown in \fref{fig:baffleinstall}, where a simple shielding of the exposed entrance aperture reduces the background significantly.
When also considering additional delay, the light scattered off beam optics can illuminate an object inside or close to the observation cone, increasing the background signal at later times, \revision{added example}{for example giving rise to the peak at $t=\SI{2.7}{\nano\second}$.}
With the same argument as above, light preceding the main pulse could therefore contribute to the background signal around $t=0$ via the delayed paths.
On this shorter time scale, the contribution of ASE or pre-pulses may not be negligible.
\revision{Refined numbers with more sensitive measurement}{The JETi-200 laser exhibits a temporal intensity contrast better than {\num{1e-10}} at times earlier than {\SI{50}{\pico\second}} before the main pulse.}
Integrating the peak intensity over \SI{24}{\femto\second} and the ASE intensity level over the gating time of \SI{1}{\nano\second} yields a ratio of \num{4e-6} in terms of energy/ photon number.
As the main pulse contributes $\approx 6300\perJperROI$ photons at $t=\SI{2.7}{\nano\second}$, the ASE could therefore contribute $0.03\perJperROI$ photons at $t=0$.
For a prepulse to contribute 1 photon per Joule at $t=0$, it would have to be at a power fraction of $1/6300\approx\num{2e-4}$ compared to the main pulse.
Typical high power laser systems exhibit prepulses at lower than \num{1e-6} to \num{1e-8} level on the picosecond to nanosecond time scale, the contribution from prepulses can thus be neglected.
Finally, light arriving at later times can still contribute to the signal at $t=0$ if the detector does not turn off completely, i.e. the high voltage on the image intensifying micro-channel plate takes some time to return to \SI{0}{\volt}.
\revision{Added measurement}{This was tested in a separate setup by shining attenuated pulses directly onto the camera and comparing the number of photons registered for times} \SIrange{1}{3}{\nano\second} before $t=0$ to those registered around $t=0$. This results in an upper bound for the gating suppression ratio of $\num{1e-4}$ at \SIrange{1}{3}{\nano\second} before $t=0$.
Light of the peak around $t=\SI{2.7}{\nano\second}$ ($\approx 6300\perJperROI$) can therefore contribute no more than $0.6\perJperROI$ photons at $t=0$ via camera \enquote{leakage}.

\revision{Added for clarity}{Taken together, these temporal effects can therefore explain no more than $\approx0.6\perJperROI$ photons recorded at $t=0$.}
The majority of the pressure-independent background contribution must originate from directly scattered light of the main pulse, arriving within $\ll\SI{1}{\nano\second}$ with respect to the pressure-dependent signal from the rest gas scattering.
Primarily, this is expected to originate from light scattered directly off the focusing optic.
However, any upstream optic can also contribute, since light scattered off these optics is concentrated by the focusing optic.
This can be easily understood by tracing the rays of a point source imaged by the OAP to a finite imaging distance behind the focus.
The closest upstream optic is the $\lambda/2$-plate (only present for measurements at vertical polarization) at a distance to the OAP of $\approx\SI{30}{\centi\meter}$.
A real image with $1.5\times$ magnification is then formed at \SI{45}{\centi\meter} behind the OAP.
Part of the rays forming the image intersect the opening aperture of the collection optic, hence light scattered off the $\lambda/2$-plate can contribute significantly to the background signal.
For an object distance $>\SI{61}{\centi\meter}$ no more imaging rays intersect the opening aperture at the right angle, so optics further upstream will have a smaller impact.

\revision{reworded for clarity}{In summary, assuming other scattering sources (e.g. optics $\gg\SI{61}{\centi\meter}$ upstream) are much less significant than the OAP and the $\lambda/2$-plate, the total number of observed background photons can be split between the contributions identified.
As all effects are based on scattering from imperfect surfaces, no significant polarization-dependence of each contribution is expected and difference between the total scattered light for horizontal and vertical polarization can be attributed to the presence of the $\lambda/2$-plate.}
\Tref{tab:numbersummary} summarizes this result.
\begin{table}
	\caption{\label{tab:numbersummary}
	Summary of estimated pressure-independent background photons per \SI{1}{\joule} per \SI{100}{\micro\meter\squared}
	in focus attributed to different sources. \revisionfig{reordered for clarity, numbers refined see text.}}
	\begin{indented}
		\item[]\begin{tabular}{@{}lll}
			\br
			Polarization       & Horizontal & Vertical \\
			\mr
			Total              &     43     &     62 \\
			\mr
			ASE                &    0.03     &    0.03 \\
			Gating suppression &    0.6      &     0.6 \\
			OAP                &     42     &     42 \\
			$\lambda/2$-plate  &   n.a.     &     19 \\
			\br
		\end{tabular}
	\end{indented}
\end{table}

\section{Extrapolation to Petawatt scale}
\label{sec:pw_scale}

\begin{figure}
    \begin{center}
	    \includegraphics[width=.8\textwidth]{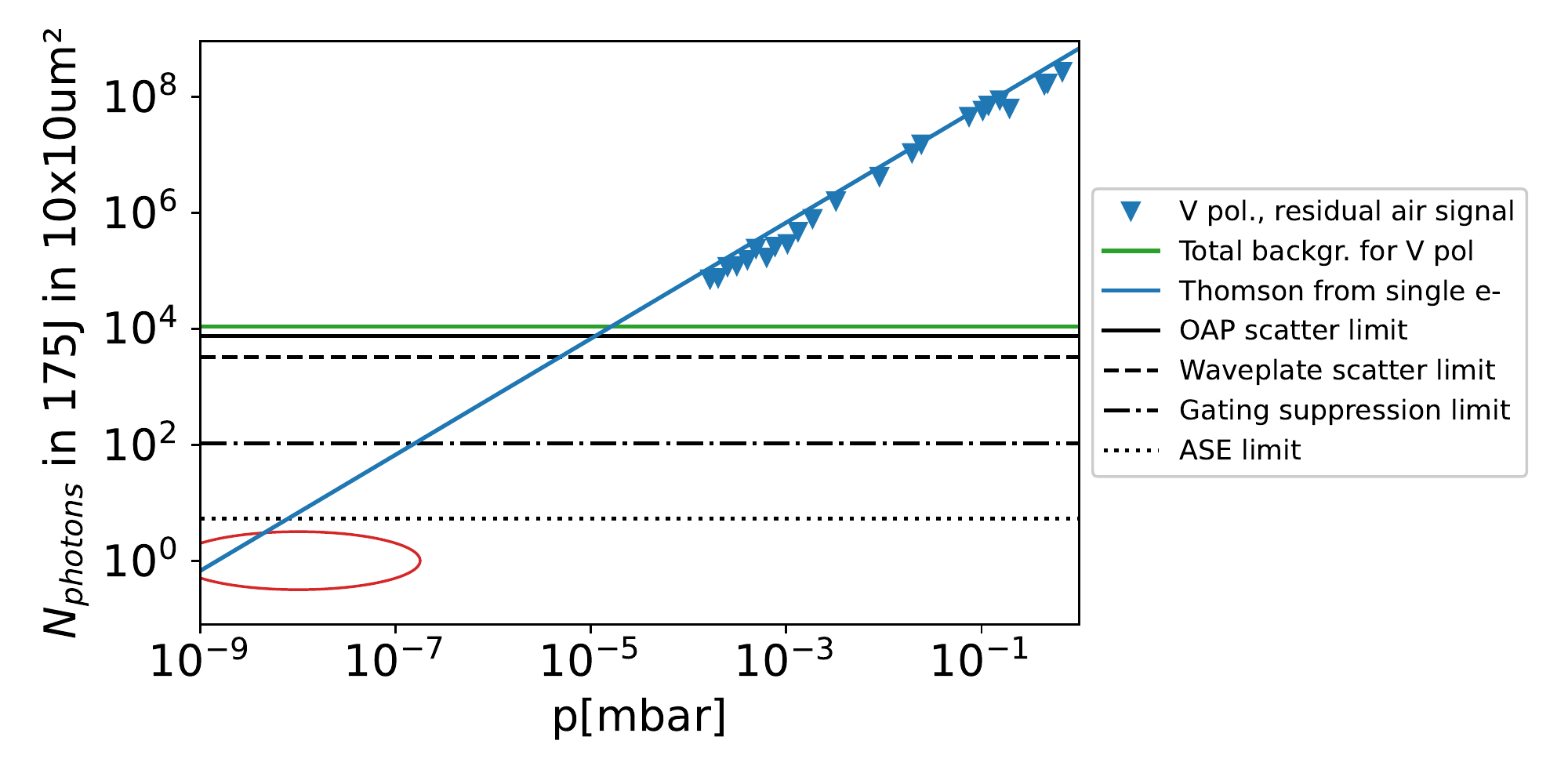}
	\end{center}
	\caption[Scaled estimate for the PW level]{\label{fig:pscan_scaled}
	\revisionfig{Tidied up, added ellipse for visual guide.}
	Simple estimate for the expected background signal contributions at \SI{175}{\joule} laser energy ($\approx\SI{5}{\peta\watt}$). The out-of-cone background was subtracted from the focus ROI data to show pressure-dependence more clearly. To reach the single photon level (red ellipse), a reduction of scattering by $10^3\times$ is necessary for the given setup, while at a pressure of \mbar{1e-7} below 100 photons are expected, potentially reduced further by electron cavitation due to higher peak intensities.}
\end{figure}

\revision{Added for context}{To provide insights and potential guidelines for future photon-photon scattering experiments, the results need to be put into context of higher laser peak power and potentially improved detection schemes.
To this end, the recorded data is scaled to {\SI{175}{\joule}} pulse energy and the pressure-independent background of the out-of-cone ROI is subtracted from the signal in the focus ROI.
Hereby we can provide estimates for the parameters required to reach single photon level inside the focal ROI.}
This corresponds to the conditions with a peak power of $\approx\SI{5}{\peta\watt}$ as expected e.g. at the \enquote{Extreme Light Infrastructure} (ELI)\cite{turcuHighFieldPhysics2016}.
\Fref{fig:pscan_scaled} visualizes the estimated background contributions for these parameters.
At a residual gas pressure on the order of \mbar{1e-7} the rest gas will contribute $<100$ photons per \SI{100}{\micro\meter\squared} on average assuming linear scaling. The ponderomotive electron cavitation will play a vital role to reduce scattering from the focal volume further and might make other (electron) cleaning methods obsolete.
\revision{Reworded to avoid repetition}{While the scattering cross sections of the nuclei will remain negligible, }
the electric fields of these nuclei could affect the quantum vacuum state and hence one needs to consider removing them dynamically.

\revision{added text to support graph}{Assuming similar components and detectors are used as in this setup, the static background would have to be reduced by 3-4 orders of magnitude to reach the single photon level.}
By shortening the gating time from \SI{1}{\nano\second} to the pulse duration scale, e.g. based on optical switching, the ASE contribution can be suppressed by $\frac{\SI{1}{ns}}{\SI{24}{fs}}=\num{4e4}$ times to a negligible level.
The gating suppression of the detector discussed will also be improved in a similar order.
For a prepulse to contribute to the signal at $t=0$, it would have to exceed an energy level of $1/(\SI{6300}{\per\joule}*\SI{175}{J})\approx\num{1e-6}$ of the main pulse as well as matching additional path length from a scattering surface.
For typical high power laser systems, prepulses can be suppressed well below this level.
Photons scattered from optics likely remain the strongest contribution to the background and must be reduced by using surfaces with (much) higher optical quality. The static-scattering improvement will need to be determined by future experiments.

Most critical is the aspect that a realistic quantum vacuum experiment will require at least two beams and the few signal photons will not emerge in \ang{90}, therefore the collection optic will point more towards one of the focusing optics.
In this case we expect a rapid, yet to be determined, growth of the static scattering background contribution.
Shortening the gating time of the detector from ns to ps or even fs therefore seems necessary to discriminate these photons.

\section{Conclusion}
\label{sec:conclusion}

We have investigated scattering from \SI{175}{\milli\joule} laser pulses tightly focused in the residual gas environment of an interaction chamber. Scaling to 1000 times larger pulse energy reveals that the residual gas density in the focus is likely not the limiting factor, in particular due to the ponderomotive cavitation.
Significant improvement of surface quality of involved optics and reduction of temporal gate times from nanoseconds to femtoseconds are key to expedient experiments. Ideally, those technical improvements can yield a discrimination of static scatter by 5-6 orders of magnitude.
Finally, an ideal setup should avoid all direct light paths from single scattering events into the collection optics of the signal chain.

\section{Acknowledgements}

The authors wish to thank all technicians and colleagues at the JETi laser for their support during the experimental campaign.
This work has been funded by the Deutsche Forschungsgemeinschaft (DFG) under Grant No. 416702141 within the Research Unit FOR2783/1.

\printbibliography

\end{document}